\documentclass[12pt]{article}

\usepackage{scicite,graphicx}

\usepackage{times}

\usepackage{amssymb,amsmath,bm}

\newenvironment{sciabstract}{%
\begin{quote} \bf}
{\end{quote}}

\title{Simulations predict intermediate-mass black hole formation in globular clusters}

\author
{Michiko S. Fujii$^{1\ast}$, Long Wang$^{2}$, Ataru Tanikawa$^{3}$, Yutaka Hirai$^{4,5}$, 
Takayuki R. Saitoh$^6$}

\date{}

\newcommand{\apjl}{Astrophys. J. L.}
\newcommand{\apjs}{Astrophys. J. S.}

\newcommand{\aap}{Astron. Astrophys.}

\begin{document} 


\baselineskip24pt


\maketitle 
\normalsize{$^{1}$Department of Astronomy, The University of Tokyo, Bunkyo-ku, Tokyo 113-0033, Japan}\\

\normalsize{$^{2}$ School of Physics and Astronomy, Sun Yat-sen University, Zhuhai 519082, China}\\

\normalsize{$^{3}$Center for Information Science, Fukui Prefectural University, Eiheiji-cho, Fukui 910-1142, Japan}\\

\normalsize{$^{4}$Department of Physics and Astronomy, University of Notre Dame, Notre Dame, IN 46556, USA}\\

\normalsize{$^{5}$Astronomical Institute, Tohoku University, Sendai, Miyagi 980-8578, Japan}\\

\normalsize{$^{6}$Department of Planetology, Kobe University, Kobe, Hyogo 657-8501, Japan}\\
\\
\normalsize{$^\ast$ Corresponding author. E-mail:  fujii@astron.s.u-tokyo.ac.jp.}

\begin{sciabstract} %
Intermediate-mass black holes (IMBHs) are those between 100 and 10$^5$ solar masses ($M_{\odot}$); their formation process is debated. 
One possible origin is the growth of less massive black holes (BHs) via mergers with stars and compact objects within globular clusters (GCs). However, previous simulations have indicated that this process only produces IMBHs $<500 M_{\odot}$ because the gravitational wave recoil ejects them when they merge with other BHs. 
We perform star-by-star simulations of GC formation, finding that high-density star formation in a GC's parent giant molecular cloud can produce sufficient mergers of massive stars to overcome that mass threshold.
We conclude that GCs can form with IMBHs $\gtrsim 10^3 M_{\odot}$, which is sufficiently massive to be retained within the GC even with the expected gravitational wave recoil.
\end{sciabstract}

\clearpage
Intermediate-mass black holes (IMBHs) have masses in the range $100$ to $10^5 M_{\odot}$ {between those of} stellar-mass ($<100 M_{\odot}$) and supermassive ($>10^5M_{\odot}$) {black holes (BHs)}.
Stellar-mass BHs {are observed in interacting} binar{y star systems}\cite{2016A&A...587A..61C}, and gravitational waves {generated by their mergers} \cite{2016PhRvX...6d1015A}. Stellar-mass BHs up to $\sim 50 M_{\odot}$ {are formed in supernova explosions of massive stars, in abundances that are consistent with} stellar evolution models\cite{2017ApJ...836..244W}. 
Supermassive BHs {are observed} at the centers of galaxies\cite{2013ARA&A..51..511K} {including the Milky Way} \cite{2019A&A...625L..10G}.
{Stellar and supermassive BHs have been confirmed by multiple lines of observational evidence; in contrast,} the existence of IMBHs {is not well established and subject to debate}\cite{2002AJ....124.3270G,2003ApJ...582L..21B,2005ApJ...634.1093G,2017MNRAS.464.2174B}.
{At the low mass end of the IMBH range, the observation of gravitational wave event GW 190521 has indicated} the formation of a $146 M_{\odot}$ BH {in} a merger of {two BHs with masses of} 85 and $66 M_{\odot}$ \cite{2020PhRvL.125j1102A}.  
{Theoretical analyses have predicted that such merger events most often occur} in star clusters{, where the density of stars and BHs is high} \cite{2015MNRAS.454.3150G, 2021MNRAS.501.5257R, 2021MNRAS.507.5132D,2022ApJ...940..131G}. 
{At the high mass end of the IMBH range, observations of} active galactic nuclei in dwarf galaxies {have indicated the presence} of IMBHs {with masses} $10^4$ {to} $10^5M_{\odot}$ \cite{2017IJMPD..2630021M}.

\section*{IMBHs in globular clusters}
Globular clusters (GCs) are {gravitationally bound} star clusters with {masses of} $10^5$ {to} $10^6 M_{\odot}${. They consist of old stellar populations and are} located in the halo of the Milky Way {and other galaxies.}  
Dynamical models and velocity measurements of {some} GCs {have indicated the possible presence of} IMBHs with masses $10^3$ {to} $10^4M_{\odot}$ at their centers\cite{2013A&A...552A..49L}. However, velocity distribution can also be explained by {dense} concentrations of stellar-mass BHs and neutron stars {at GC centers} \cite{2003ApJ...582L..21B,2020ApJ...898..162W}.

{Theoretical studies have difficulty explaining the formation of} $500$ {to} $10^4 M_{\odot}$ IMBHs in GCs, if they exist. 
In the {center of GCs, their most dense region,} stars and BHs can collide and merge{, potentially producing} IMBHs\cite{2002ApJ...576..899P,2015MNRAS.454.3150G,2022ApJ...927..231F}.
{Numerical simulations have predicted that} mergers of binary BHs {can form} IMBHs up to $\sim 500 M_{\odot}$ \cite{2021MNRAS.501.5257R,2021MNRAS.507.5132D,2022ApJ...940..131G}. However, {BH mergers emit} gravitational waves{; if the masses or spins of the BHs differ from each other, the emission is asymmetric, imparting a recoil kick velocity which can eject the resulting IMBH from its host cluster}\cite{2008ApJ...686..829H,2022ApJ...940..131G}. This ejection mechanism {prevents low-mass IMBHs from growing to more massive IMBHs via BH-IMBH mergers in GCs.}

{An alternative scenario for IMBH formation is very massive stars (VMSs). Their masses are typically $>10^3 M_{\odot}$, which is much more than those of any currently observed stars that are limited to $\sim 200 M_{\odot}$ \cite{2013A&A...558A.134D}.
VMSs are considered to be formed via multiple mergers of massive stars (runaway collision) before they evolve into BHs \cite{2002ApJ...576..899P}. Theoretical studies have suggested that VMSs finally collapse to IMBH \cite{2002ApJ...567..532H}. This IMBH formation scenario occurs only in extremely dense stellar systems, such as the central regions of {GCs} \cite{2002ApJ...576..899P,2004Natur.428..724P}. Although numerical studies showed the successful formation of VMSs via runaway collisions\cite{2002ApJ...576..899P,2004Natur.428..724P}, later studies demonstrated that VMSs in GCs rapidly lose their mass through stellar winds, so evolve into BHs with masses considerably lower than $10^3 M_{\odot}$. The mass lost through winds is expected to depend on the VMSs' metallicity (abundance of elements heavier than helium); those simulations assumed the VMS has a similar metallicity to GCs observed in the Milky Way \cite{2009A&A...497..255G}.}

The {maximum} VMS mass {results from a} competition between the rate {of mass accumulation from stellar mergers} and the mass-loss rate {from stellar winds} \cite{2009A&A...497..255G}. 
{Previous simulations assume initial conditions of GCs that were} gas-free and in dynamical equilibrium. {In those simulations, runaway collisions started after the dynamical evolution time of GCs}, and the {mass accumulation rate} of VMSs decreased with time~\cite{2009ApJ...695.1421F} {because all massive stars initially in the GCs merged into the VMSs}. 
{Those simulations also found that} the density of isolated clusters decreased {over time, due} to dynamical scattering of stars in the cluster center and mass loss due to stellar {winds}~\cite{2010ARA&A..48..431P}. {We propose these are} inevitable {consequences when simulating} star clusters in which star formation has already ended.

{Those previous simulations have oversimplified the initial condition of GCs. GCs are not initially gas-free nor in dynamical equilibrium. Stars in star clusters are generally formed from gas and dust in giant molecular clouds (GMCs)~\cite{2003ARA&A..41...57L}. }
In this phase, {young} star clusters are embedded in their {natal} GMCs, and {more of the mass is in gas than in stars}. {The additional gravitational potential provided by the gas keeps the cluster compact while it forms. This increases} the stellar density {and therefore the probability of stellar mergers to the VMSs,} until gas {has been completely ejected from the GC (gas expulsion) due to the radiation from massive stars~\cite{2018NatAs...2..725H}.} 
{During this gas-rich phase, we anticipate that continuing star formation provides an ongoing supply of massive stars, which could potentially merge with the VMS. 
However, no simulations have considered both stellar collisions forming a VMS and ongoing star formation due to the computational cost of doing so.}

\section*{Simulations of globular cluster formation}
{We simulated GC formation, including star formation and mergers producing VMSs. The integration scheme we adopt~\cite{2020MNRAS.497..536W,2021PASJ...73.1074F} allows our simulations to resolve the accurate motions of individual stars in the GC.}
{To reduce the computational} cost{, previous simulations used} star particles {that} did not represent individual stars {but a group of stars}~\cite{2018NatAs...2..725H,2021MNRAS.506.5512F}. However, the formation of VMSs via runaway collisions {requires} resolving individual stars {in the simulations (hereafter referred to as star-by-star simulations), and the} integration of stellar orbits {during} the rapid {changes in orbit immediately prior to} collisions. Our {simulation }code \cite{supplementary} {incorporates an integration scheme that handles} such close encounters and {stellar }binaries \cite{2020MNRAS.497..536W}.

{Our initial conditions are GMCs with masses of $10^5$ and $10^6M_{\odot}$, consisting of gas with} turbulent velocity fields {and} a metallicity $Z$ similar to that of GCs ($Z=0.02\,Z_{\odot}${, where $Z_{\odot}$ is the metallicity of the Sun}){. In the resulting simulations, the} clumpy and filamentary structures of the {GMCs lead to clumpy star formation} (Fig.~1A). These stellar clumps merged to form a massive cluster. At typically 0.7 {to} 0.8\,Myr (Table S3), the gas was completely ejected from the cluster, {ending} star formation. The {cluster produced in the simulation with $10^6M_{\odot}$ reached a maximum gravitationally bound mass} of $\sim 4\times 10^5M_{\odot}$, {which is approximately} the mass of {a} small GC \cite{2018MNRAS.478.1520B}. 

\begin{figure}[h]%
\centering
\includegraphics[width=1.\textwidth]{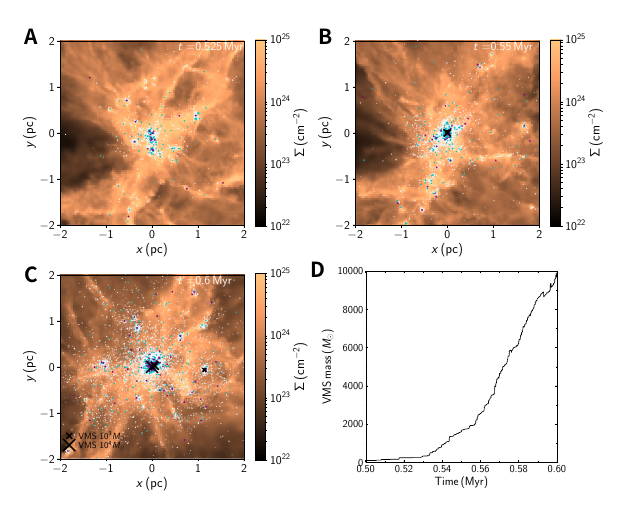}
\caption{{\bf Time evolution of an example of the simulations.} (A) {to} (C) Snapshots of the star cluster formed in the simulation {(model M1e6R10Z002v01s2 in Table S2) at 0.525, 0.55, and 0.6\,Myr from the beginning of the simulation. The $x$-$y$ coordinate shows the distance centered by the most massive star in the cluster.} The color scale indicates the {gas column number} density ($\Sigma$), and the dots indicate stars {formed in the simulation (violet: $m\geq40M_{\odot}$, blue: $10\leq m<40M_{\odot}$, and white: $m<10M_{\odot}$).} Black crosses indicate VMSs with $m>300M_{\odot}$.  
(D) Time evolution of the most massive VMS mass  of model M1e6R10Z002v01s2.}\label{fig1}
\end{figure}

{Within each GC formed in our} simulations, a VMS was also formed via runaway collisions. {Fig.~\ref{fig1}D shows the} {time evolution of the mass for an example} VMS {taken from the most massive} one of our simulations. The {mass accumulation rates} of the VMSs {in our simulations peaked at} $\sim 0.01$ {to} $0.1 M_{\odot}$\,yr$^{-1}$, {equivalent to} $\sim 3$\,\% of the star formation rate of the cluster (Fig.~S10).
These {mass accumulation rate}s are larger than the mass-loss rate of $10^{-3}M_{\odot}$\,yr$^{-1}$ (by stellar winds) for stars with a mass of $\sim10^3 M_{\odot}${. Mass loss rates ($\dot{m}_{\rm SW}$) were determined }from {the trend found in a} stellar evolution model for {VMSs in the range} $300$ {to} $1000 M_{\odot}$:
\begin{equation}
\log[\dot{m}_{\rm SW}/M_{\odot}\,{\rm yr}^{-1}] = -9.13+2.1\log[m/M_{\odot}] + 0.74\log[Z/Z_{\odot}],\label{eq:Vink}
\end{equation}
where $m$ is the current mass of the VMS \cite{2018A&A...615A.119V}.
{Our} simulations{ were} terminated just before complete gas expulsion for clusters with $>10^5M_{\odot}$ and after gas expulsion for less massive clusters. After the gas expulsion, {we assumed that each} VMS {continued to lose} mass according to equation~(\ref{eq:Vink}). We estimated the final remnant {IMBH} mass using a stellar evolution {simulation} code~\cite{2002MNRAS.329..897H, 2020A&A...639A..41B}.
The {simulated final} cluster and IMBH masses are shown in Fig.~\ref{fig2}. {All our simulations that produced} star clusters with $>10^5M_{\odot}$ resulted in the formation of {an }IMBH with $>10^3M_{\odot}$. 

{We performed }additional simulations {of GMCs with} $10^5M_{\odot}$ {and} a higher metallicity ($Z=0.1Z_{\odot}$); the results are also shown in Fig.~\ref{fig2}. 
{The lower GMC masses produced} $\sim 10^4M_{\odot}$ star clusters{; in this case we found that} the metallicity {does} not affect the IMBH mass because the mass-loss rate {remains} much lower than the {mass accumulation rate}. For more massive star clusters, {the higher metallicity leads to lower} IMBH mass because the higher mass-loss rate {is sufficient to overcome the mass gained through mergers.}

\section*{Theoretically expected IMBH masses}
{In all our simulations}, the {mass accumulation rate} of VMSs due to runaway collisions ($\dot{m}_{\rm col}$) was $\sim 3$\,\% of the star formation {rate of the cluster} ($\dot{M}_{\rm SF}$) regardless of the {assumed }metallicity (Fig.~S10). This {scaling is consistent with} a theoretical model for the collisions of {stars via stellar binaries} in star clusters \cite{2002ApJ...576..899P}.
The {mass accumulation rate} due to collisions {is}:
\begin{eqnarray}
    \dot{m}_{\rm col} = \dot{N}_{\rm col} \langle\delta m\rangle,\label{eq:dmdt}
\end{eqnarray}
where $\dot{N}_{\rm col}$ and $\langle\delta m\rangle$ are the collision rate and the average mass {of the colliding stars}, respectively.
The collision rate {is related to} the {stellar }binary formation rate via three-body encounters {and collision fraction of binaries} as:
\begin{eqnarray}
    \dot{N}_{\rm col} \sim 10^{-3} f_{\rm c}\frac{N_{\rm cl}}{t_{\rm rlx}},\label{eq:collision_rate}
\end{eqnarray}
where $f_{\rm c}\,(0<f_{\rm c}<1)$ is the fraction of binaries {colliding after the formation}, $N_{\rm cl}$ {is the number of stars in the cluster,} and $t_{\rm rlx}$ {is} the half-mass relaxation time of the cluster. {The half-mass relaxation time is a timescale with which star clusters dynamically evolve; for example, massive stars sink to the cluster center on this timescale by giving the kinetic energy to the less massive stars in the star clusters.} The average mass of merging stars $\langle \delta m\rangle$ is {related} time {$t$ as}: 
\begin{eqnarray}
    \langle\delta m\rangle \simeq 4\frac{t_{\rm rlx}}{t}\langle m\rangle \ln \Lambda,\label{eq:dm}
\end{eqnarray}
where $\langle m\rangle$ is the mean stellar mass {in the cluster}, which is determined from the {assumed }initial mass function{ of the star formation}, and the factor {of 4 has been} empirically determined \cite{2002ApJ...576..899P}. Here, $\ln \Lambda$ is called the Coulomb logarithm, and the value depends on the distribution of stars. Substituting equations (\ref{eq:collision_rate}) and (\ref{eq:dm}) into equation (\ref{eq:dmdt}), we obtain
\begin{equation}
    \dot{m}_{\rm col} \sim 4\times 10^{-3} f_{\rm c} \ln \Lambda \frac{N_{\rm cl}\langle m\rangle}{t},\label{eq:dmdt_mid}
\end{equation}
where $N_{\rm cl} = M_{\rm cl}(t)/\langle m\rangle$ {is now} a function of time because the cluster mass increases {over} time. For the Coulomb logarithm, we adopt $\ln \Lambda = \ln (0.1 N_{\rm cl})$\cite{2003gmbp.book.....H}, {so} $\ln \Lambda \sim 7$ {to} $12$ for $N_{\rm cl}=10^4$ {to} $10^6$. With these {values}, we obtain
\begin{equation}
    \dot{m}_{\rm col} \sim 0.04 f_{\rm c} \dot{M}_{\rm SF}.\label{eq:dmdt_final}
\end{equation}
Adopting $f_{\rm c}=0.8$ {following a previous study considering the existence of gas in star clusters}\cite{2022MNRAS.512.6192S}, we obtain $\dot{m}_{\rm col} \sim 0.03 \dot{M}_{\rm SF}${, identical to the $\sim$3\% that appears in our simulations}. {Equation (\ref{eq:dmdt_final}) demonstrates} that the {mass accumulation rate} due to runaway collisions does not {directly} depend on the cluster mass but on the star-formation rate of the cluster.

High star-formation and collision rates {occur due to} the high density of the forming star clusters{ in our simulations}. {Figs.~\ref{fig3}A and B show} the radial distribution of gas and stars {in a simulated} star cluster and the mass function of stars {that }merged into the VMS. {During} the cluster-formation phase, the gas density exceeds the stellar density {at} the cluster center, {while} star formation {occurs} in the central region of the star cluster. {In our simulations, the} stellar density exceeds $10^7M_{\odot}$\,pc$^{-3}${ at the cluster center}, which is higher than the density necessary for runaway collisions~\cite{2008ApJ...682.1195A}. {Fig.~\ref{fig3}C shows} the mass distribution of stars {that }merged into the VMS{; we find a higher proportion of massive stars merged than that produced by the assumed} initial mass function. {We interpret this as the result of mass segregation, which preferentially moves more massive stars towards the cluster center. As a result, the} mass function of {surviving stars in} the cluster {has fewer massive stars than were initially formed, due to the loss of those which merged with the VMS or were ejected from the cluster}~\cite{2011Sci...334.1380F}.

\section*{Comparison with Milky Way GCs}
{Fig.~\ref{fig2} compares the} IMBH masses {predicted by our simulations with observational estimates of possible IMBHs} in some GCs {in the Milky Way}. {Our simulations that used 10$^6$ $M_{\odot}$ GCs produced} IMBH masses {similar} to those {that have been} observationally estimated. Milky Way GCs {have aged for billions of years, during which they have lost mass due} to dynamical evolution, stellar evolution, and tidal disruption\cite{2003MNRAS.340..227B,2011MNRAS.413.2509G,2015MNRAS.453.3278W}. {Previous work has estimated that} the current {masses of Milky Way GCs can be} $\sim 10$\,\% of {their} initial mass{es at formation} \cite{2015MNRAS.453.3278W}.
{If we assume} the Milky Way GCs {had initial masses} $10^6$ to $10^7 M_{\odot}$, we expect {those would correspond to} IMBH masses {up to} $10^4M_{\odot}$ because the mass-loss rate of VMSs increases with $\propto m^{2.1}$ ({equation~\ref{eq:Vink}), and we determined above that IBMHs in high mass GCs are limited by the VMS stellar wind. In contrast}, the {mass accumulation rate} {is linearly proportional to the} cluster mass ({equation~\ref{eq:dmdt_mid}}). {We use semi-analytic methods \cite{supplementary} to} estimate the mass of IMBHs {we expect to form} in GCs as a function of the cluster mass and metallicity{; the resulting predicted relation is also} shown in Fig.~\ref{fig2}. {In this calculation, the} IMBH mass is an upper limit because {the semi-analytic method assumes} that clusters form in less than 1\,Myr. {The predicted relations flatten towards higher cluster masses, so we conclude that} the IMBH mass {produced by the VMS} cannot be larger than $10^4 M_{\odot}${,} even if the host cluster {is very} massive. 

{We also compare the rotation rate} of clusters in simulations {to observed} GCs. 
Some Milky Way GCs are known to rotate~\cite{2018MNRAS.481.2125B}, and {some previous simulations of small clusters have produced} cluster rotation \cite{2017MNRAS.467.3255M,2020MNRAS.496...49B}.
We {determined} the cluster rotation {rates} of the simulated star clusters\cite{supplementary} and compare them {to observed} Milky Way GCs in Fig.~\ref{fig:rotation}. The {simulated GCs} initially rotate, {and those with $>10^5M_{\odot}$ rotate faster than the typical rotation speed of the Milky Way GCs. This result matches the scenario that }the rotation slow{s} down as the clusters dynamically evolve and are tidally disrupted \cite{2017MNRAS.469..683T}.

\section*{The fate of IMBHs in GCs}
\begin{figure}[h]%
\centering
\includegraphics[width=0.9\textwidth]{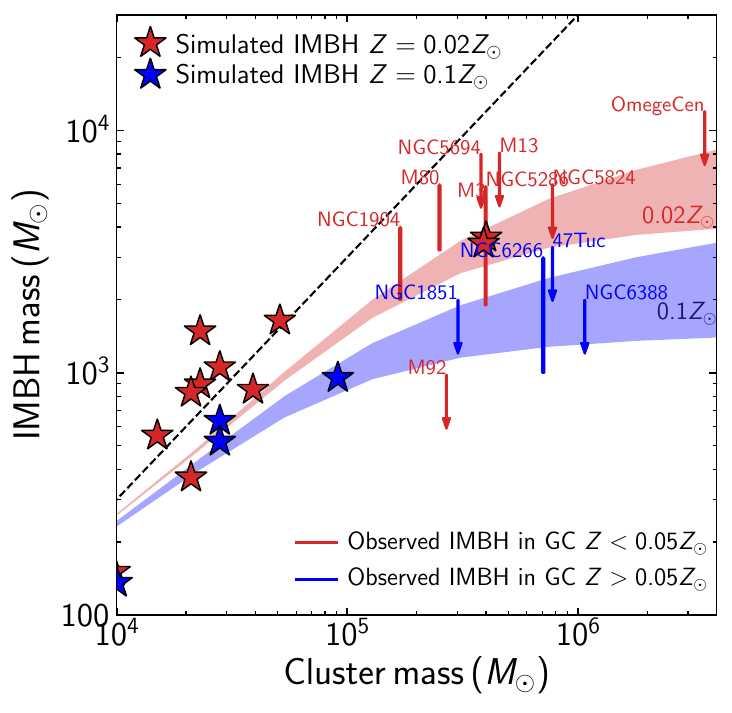}
\caption{\textbf{Simulated and observed IMBH masses in GCs.}
Star symbols {are} the results of {each of }our simulations. 
The dashed line {is} the relation
{that IMBH mass is 3\,\% of cluster mass, predicted by equation \ref{eq:dmdt_final}}. 
The {colored }lines and arrows ({the latter indicating} upper limit{s}) {are} observationally estimated IMBH masses {for} Milky Way GCs ({labels correspond to entries in} Table S4). 
{Red and blue are different metallicity ranges (for the observations) or assumptions (for the simulations), as indicated by the legends.} The colored regions are the upper-mass limits of IMBHs estimated from {our semi-analytic calculation} based on the simulation results {and theoretical estimation}~\cite{supplementary}.
}\label{fig2}
\end{figure}

\begin{figure}[h]%
\centering
\includegraphics[width=0.9\textwidth]{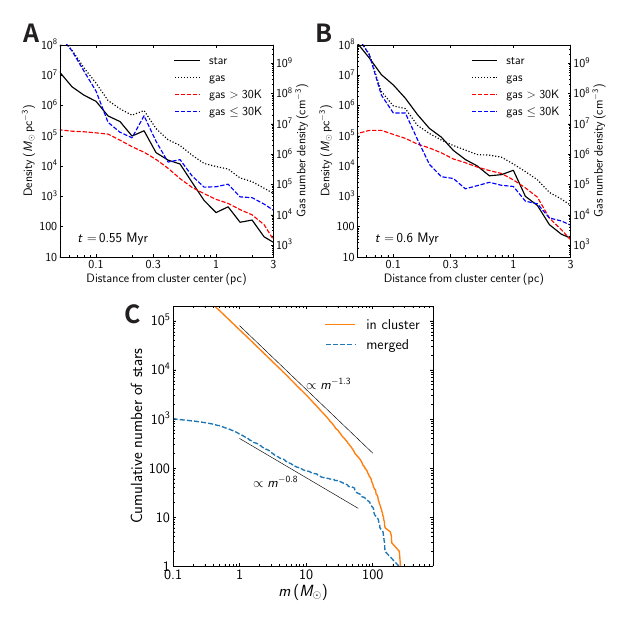}
\caption{\textbf{Structures of the simulated cluster.} (A) to (B) Density distribution of stars ({solid }black {line}) and {gas (red for gas $>$30 K, blue for gas $\leq$30 K, and dotted black for their total) as a function of the distance from the center cluster} at $t=0.55$ and 0.6\,Myr for model M1e6R10Z002v01s2. {The left and right axes are the density in $M_{\odot}\,{\rm pc}^{-3}$ and hydrogen number density for gas in ${\rm cm}^{-3}$, respectively.} (C) Cumulative mass distribution of stars {that} merged into IMBHs (blue dashed curve) and in the {rest of the} cluster (orange full curve) at $t=0.6$\,Myr for model M1e6R10Z002v01s2. The {black} line {labeled} $\propto m^{-1.3}$ indicates the assumed initial mass {function}. The {black }line {labeled} $\propto m^{-0.8}$ indicates {a} power-law {estimated from} the merged stars. 
}\label{fig3}
\end{figure}

\begin{figure}[h]%
\centering
\includegraphics[width=0.9\textwidth]{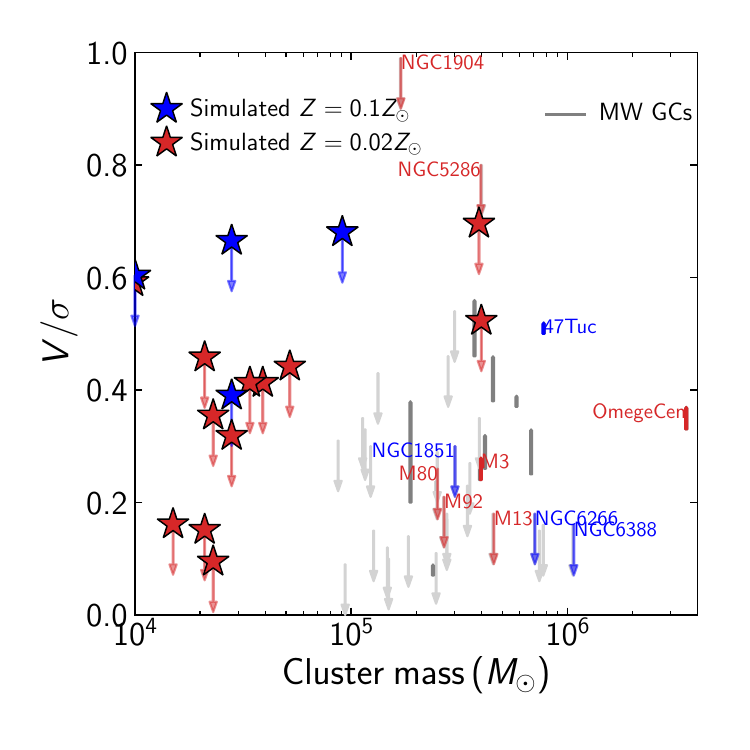}
\caption{\textbf{Cluster rotation velocity as a function of GC mass.} Cluster rotation velocity ($V$) is scaled by one-dimensional velocity dispersion ($\sigma$). Star symbols indicate the maximum rotation at the gas expulsion time{ in our simulations}. Gray indicates {rotation rates} of Milky Way GCs \cite{2018MNRAS.481.2125B}{; dark gray lines are measured values, while light gray arrows are upper limits.} Colored lines and arrows {with labels are}
the {same observed }GCs {as }shown in Fig.~\ref{fig2}. {Red and blue have the same meaning as in Fig.~\ref{fig2}, for both the simulations and observations.}
}\label{fig:rotation}
\end{figure}

{Our simulations indicate that GCs can form IMBHs at their centers during the star formation process. We next consider whether IMBHs could still be present in Milky Way GCs.}
Some previous studies\cite{2008ApJ...686..829H,2022ApJ...940..131G} have {found} that IMBHs {are} ejected from {their} host clusters when they merge with another BH{, due} to asymmetric gravitational wave emission {(as discussed above).}
The recoil velocity {from such a merger }depends on the spin {of both BHs and their masses. Previous studies found that the probability of IMBHs retained within a GC after 25 mergers with a realistic mass distribution of BHs falls from $\sim80$ to $\sim30$\,\% as the IMBH mass decreases from 2000 to $1000M_{\odot}$ and almost 0\,\% for $<500M_{\odot}$\cite{2008ApJ...686..829H}.} {If Milky Way GCs have lost more than half of their initial masses as previous studies suggested (ref), their typical initial mass is $\gtrsim10^6M_{\odot}$. In this cluster mass range, our simulations suggest that star clusters with metallicity of $<0.1 Z_{\odot}$ can initially host IMBHs with $\gtrsim 2000 M_{\odot}$ (Fig.~\ref{fig2}). Such sufficiently massive IMBHs cannot be ejected from their host clusters even after multiple mergers with other BHs. On the other hand, high-metallicity ($>0.1Z_{\odot}$) GCs, which are one-third of the Milky Way GCs\cite{1996AJ....112.1487H,2010arXiv1012.3224H}, would initially have hosted less massive IMBHs and have lost them.} 
Observations {have shown} that {at least some} GCs do not host IMBHs\cite{2003ApJ...582L..21B}.
{These GCs may have lost their IMBHs by the ejection due to IMBH-BH mergers. Thus, not all but a fraction of GCs, especially low-metallicity ones, still host IMBHs. }

\bibliography{scibib}

\bibliographystyle{Science}

\section*{Acknowledgments}
We thank Douglas Heggie and other anonymous referees for their constructive comments.
Numerical computations were performed on the Cray XC50 CPU cluster at the Center for Computational Astrophysics (CfCA) of the National Astronomical Observatory of Japan. 

{\bf Funding:} This work was supported in part by the Japan Society for the Promotion of Science (JSPS) through KAKENHI (Grant Numbers 17H06360 to M.S.F, 19K03907 and 24K07040 to A.T., 19H01933, 21K03614, and 22H01259 to M.S.F, T.R.S., and Y.H., 21H04499 and 22KJ0157 to Y.H., 21K03633 and 22K03688 to T.R.S) and Ministry of Education, Culture, Sports, Science and Technology (MEXT) (Grant Number JPMXP1020230406 to M.S.F, T.R.S, and Y.H). M.S.F was also supported by the University of Tokyo Excellent Young Researcher Program and the Initiative on Promotion of Supercomputing for Young or Women Researchers, Information Technology Center, The University of Tokyo. L.W. was supported by the National Natural Science Foundation of China through Grant numbers 21BAA00619, 12073090, and 12233013,  the one-hundred-talent project of Sun Yat-sen University, the Fundamental Research Funds for the Central Universities, Sun Yat-sen University (22hytd09).
Y.H. was supported by the JSPS Research Fellowship.

{\bf Authors contributions:}

M.S.F performed the simulations, analyzed the data, produced the figures, and led the manuscript writing. All authors contributed to the software development, scientific discussion, and manuscript writing.

{\bf Competing interests:} There are no competing interests to declare. 

{\bf Data and materials availability:} 
The simulation code, simulation outputs, and the scripts for analyses are available from https://doi.org/10.5281/zenodo.11180637
\cite{data}.

\clearpage

\section*{Supplementary materials}
Materials and Methods\\
Figs. S1 to S10\\
Tables S1 to S4\\
References \textit{(51--96)}\\

\renewcommand{\figurename}{Figure~S}
\renewcommand{\tablename}{Table~S}
\makeatletter
\def\fnum@figure{\figurename\thefigure}
\def\fnum@table{\tablename\thetable}
\makeatother
\setcounter{figure}{0}
\setcounter{table}{0}

\section*{Materials and Methods}
\section{Simulation code}
The simulations were performed using a smoothed-particle hydrodynamics (SPH) code,\\ \textsc{ASURA+BRIDGE} \cite{2021PASJ...73.1036H,2021PASJ...73.1057F,2021PASJ...73.1074F}, which integrates the orbits of stars in gas. \textsc{ASURA+BRIDGE} adopts a hybrid integration scheme, BRIDGE \cite{2007PASJ...59.1095F}, which applies a high-order integrator for collisional systems for stars, and a leapfrog integrator with a tree code\cite{1986Natur.324..446B} for gas particles.
For stars, we used the \textsc{Petar}\cite{2020ApJ...898..162W} $N$-body simulation code \cite{2020MNRAS.497..536W}. \textsc{PeTar} adopts a particle-particle particle-tree (P$^3$T) scheme \cite{2011PASJ...63..881O,2015ComAC...2....6I}, which is a hybrid integrator of fourth-order Hermite scheme and leapfrog integrator with tree scheme. \textsc{PeTar} also includes an algorithm for integrating binaries, using a slow-down algorithmic regularization scheme \cite{2020MNRAS.493.3398W}, which allows tight binaries to be integrated without gravitational softening. 
We adopt a Plummer-type softening between gas particles and between gas and star particles ($\epsilon_{\rm g}$) of $0.07$\,pc. 
Every Bridge timestep which lasts for a time ($\Delta t_{\rm B}$), stars are perturbed by the gravity of the gas. We adopt $\Delta t_{\rm B}=25$--200\,years depending on the gas density. 
For the radiative cooling and heating of the gas, we used \textsc{Cloudy} version13.05 \cite{1998PASP..110..761F,2013RMxAA..49..137F,2017RMxAA..53..385F} and set the minimum temperature of the gas to be 20\,K. 

We adopt a probabilistic star formation method often used for galaxy simulations, but we modified it for star formation by resolving individual stars \cite{2021PASJ...73.1036H}. In this scheme, stars are formed when gas particles satisfy the following criteria: threshold density ($n_{\rm{th}}$), threshold temperature ($T_{\rm{th}}$), and the divergence of the velocity. We assume $n_{\rm{th}}>10^5$\,cm$^{-3}$ and $T_{\rm{th}} < 30$\,K, and the divergence of the velocity is less than zero. For models with $Z=0.1Z_{\odot}$, we adopted $T_{\rm{th}} < 20$\,K, because the equilibrium gas temperature is lower than that for $Z=0.02Z_{\odot}$. Once these criteria are satisfied, stars are formed with an efficiency of 2\,\% per free-fall time {\bf \cite{2012ApJ...745...69K}}. The stellar mass is assigned following an assumed initial mass function, we adopt the Kroupa mass function with a mass range of 0.08--150\,$M_{\odot}$ \cite{2003ApJ...598.1076K}. We limit the forming stellar mass to half of the gas mass in a given search radius ($r_{\rm max}=0.2$\,pc). The newly born star has the center-of-mass position and velocity of the gas particles. This star formation procedure is implemented using the \textsc{CELib} software library\cite{2017AJ....153...85S}.
Because the star formation procedure reduces the mass of some gas particles, we merge such low-mass gas particles to the nearest gas particles when the mass becomes ten times smaller than its initial mass. 

Once massive stars are formed, feedback from massive stars is injected into the surrounding gas. \textsc{ASURA+BRIDGE} includes an H\textsc{ii} region model \cite{2021PASJ...73.1074F}, in which the Str\"{o}mgren radius is calculated using the local density around the massive star, and the gas temperature in the radius is set to be $10^4$\,K. Mechanical feedback due to the stellar wind is also imparted to the gas particles within the radius as radial velocities \cite{2013MNRAS.436.1836R}. If the Str\"{o}mgren radius is smaller than the distance to the nearest gas particle, we added the mechanical feedback to gas particles within the gravitational softening radius for gas particles. We adopted this feedback for stars with $>8 M_{\odot}$.

Our simulation includes stellar evolution using the Single Stellar Evolution (\textsc{SSE}) code \cite{2002MNRAS.329..897H} \cite{2020A&A...639A..41B} in \textsc{PeTar}. We use the stellar radius for stellar collisions and the stellar temperature for photon counts required for the H\textsc{ii} region. The photon counts are calculated using the OSTAR model \cite{2003ApJS..146..417L}, which is a table based on the numerical modeling of massive stars. We fit a function to the results at each metallicity in OSTAR,  function in our simulations. We assume Massive stars lose their mass following the mass evolution obtained from \textsc{SSE}. The mass loss rate of massive stars is constant in \textsc{SSE} in \textsc{PeTar} \cite{2020A&A...639A..41B}. We modified this from \textsc{SSE}'s constant value ($\dot{M}=1.5\times10^{-4}M_{\odot}$\,yr$^{-1}$) to equation 1, which gives a higher value. Equation 1 is applied to VMSs with $\gtrsim10^3M_{\odot}$. The lost mass is transferred to gas particles within the Str\"{o}mgren radius. 

Once stars approach closer than the sum of their radii, we assume that they merge without mass loss. To treat the stellar evolution using \textsc{SSE}, the initial mass and current age of the merged star must be given. We adopt the age of the primary (higher mass) star for the age of the merged star. We assume that the mass of the merged star is the sum of the initial mass of the primary star and the current mass of the secondary (lower mass) star.

For the radius and luminosity, we adopt the description in \textsc{SSE}, extrapolated to account for VMSs. However, \textsc{SSE} is not made for VMSs. We found that this extrapolation did not work successfully with stellar evolution for $>1000 M_{\odot}$. We, therefore, used the radius and luminosity of the zero-age main sequence for VMSs with $>1000 M_{\odot}$ (Figure~S1). The luminosity and mass-loss rate of VMSs are given independently. The resulting evolution of massive stars between 200 and $900M_{\odot}$ is shown in Figure~S2. All our simulations were halted in less than 1 Myr. 

\section{Initial Conditions}
We adopt a turbulent molecular cloud for our initial conditions \cite{2003MNRAS.343..413B}. We set up an initially homogeneous spherical molecular cloud with masses of either $10^5$ or $10^6M_{\odot}$ and metallicity of $0.02Z_{\odot}$, similar to the lowest metallicity observed for globular clusters \cite{1996AJ....112.1487H}. 
We adopt radii of 5 and 10\,pc for those two masses, respectively, to produce a similar initial density. As a result, in all cases, their initial free-fall time is 0.5--0.6\,Myr. We set an initial turbulent velocity field of $v\propto k^{-4}$ for the gas and set the initial virial ratio ($\alpha_{\rm vir}=|E_{\rm k}|/|E_{\rm p}|$, where $E_{\rm k}$ and $E_{\rm p}$ are the total kinetic and potential energies, respectively) of 0.5 or 0.1. With $\alpha_{\rm vir}=0.1$, the GMCs experience a collapse more monolithic than the case with $\alpha_{\rm vir}=0.5$. We adopt $\alpha_{\rm vir}=0.1$ for the model with $10^5M_{\odot}$ because the resulting cluster mass was too small to count a massive cluster with $\alpha_{\rm vir}=0.5$.
We set the initial gas-particle mass to be $0.1 M_{\odot}$. The initial temperature was set to be 30\,K.
In Table~S1, we summarize the parameters for each of our simulation runs.
For comparison, we also performed simulations for models with $Z=0.1Z_{\odot}$, similar to the highest-metallicity globular clusters in the Milky Way \cite{1996AJ....112.1487H}. For these higher metallicity runs, we adopted  $10^5$ and $3\times 10^5 M_{\odot}$ molecular clouds as initial conditions because $10^6 M_{\odot}$ was too computationally expensive.
Table S1 summarizes our initial conditions. We named our models using mass (``M''), radius (``R''), metallicity (``Z''), and virial ratio (``v'').

The initial turbulent velocity field is assumed to be a divergence-free Gaussian random field with a power spectrum $P(k)$, varying as $k^{-4}$, where $k$ represents the wave number \cite{2003MNRAS.343..413B}. The choice of random seed determines the shape of the structures that form in each simulation. We performed several runs for $10^5 M_{\odot}$ clusters. ``s'' in the model names indicate the random seed for the initial turbulence. The same number indicates the same initial turbulent velocity field. 

The divergence-free turbulent velocity field introduces some degree of rotation. We parameterized the rotation of initial molecular clouds with two parameters.
One is the rotational energy ($E_{\rm rot}$). 
Following previous methods\cite{2023ApJ...943...76M}, we calculate the rotational energy $E_{\rm rot}$ as
\begin{equation}
    E_{\rm rot} = \sum_{i} m_{i} \left(\frac{\bm{x}_i\times\bm{v}_i}{|\bm{x}_i|}\right)^2, \tag{S1}
\end{equation}
where $\bm{m}_i$, $\bm{x}_i$, and $\bm{v}_i$ are the mass, position, and velocity measured with respect to the center-of-mass position and velocity of each SPH particle. We define $\beta \equiv E_{\rm rot}/E_{\rm p}$, which is the rotation energy scaled to the potential energy. This value depends more strongly on the initial kinetic energy than the random seed of the turbulence (Table~S2). We also define $\gamma \equiv E_{\rm rot}/E_{\rm k}$, the ratio of rotational energy to the kinetic energy.

We also parameterized the angular momentum scaled to the cluster momentum ($\lambda$) using 
\begin{equation}
    \lambda = \frac{j}{\sqrt{2}RV}, \tag{S2}
\end{equation}
where $j$ is the specific angular momentum of the molecular cloud, and $V$ is the circular velocity at radius $R$ \cite{2001ApJ...555..240B}. 
We adopt the initial size of the molecular cloud ($R_{\rm g}$) and the circular velocity there as $R$ and $V$, respectively. If the molecular cloud rigidly rotates with the circular velocity at $R_{\rm g}$, we obtain $\lambda = 0.286$. 
We list the $\beta$ and $\lambda$ produced by each random seed in Table~S2.

\section{Summary of the simulation results}
Star formation in each simulation starts at 0.2--0.4\,Myr, and the feedback from massive stars terminates star formation before 1\,Myr. Star formation occurs in clumps (Figure~1). The clumps merge and evolve into more massive clumps. 

Collisions occur in some clumps, especially in the most massive clump in each simulation. Figure 1 shows the evolution of the mass of the most massive stars. Because the central density of the cluster exceeds $10^7M_{\odot}$\,pc$^{-3}$, the runaway collisions of stars occur \cite{2008ApJ...682.1195A}. In each simulation, $3$--5\,\% of the stellar mass formed in star clusters was provided to the VMS (Figure~S3).

After forming a massive star via collisions, the feedback (due to the H\textsc{ii} region) from the most massive star blows away the gas from the cluster, terminating the star formation. After the gas expulsion of the most massive star cluster, we stop the simulation and identify star clusters at the gas expulsion time ($t_{\rm GE}$) because several star clusters are formed in one simulation. For M1e6R10Z002v01 models, however, we stop the simulations before gas expulsion occurs due to the higher computational cost. 

We use the \textsc{HOP} density-based clump finding algorithm \cite{1998ApJ...498..137E}, as implemented in the Astrophysical Multipurpose Software Environment (\textsc{AMUSE}) software \cite{2009NewA...14..369P,2013CoPhC.183..456P, 2013A&A...557A..84P, AMUSE}. 
We use the same parameter set for \textsc{HOP} as in previous work \cite{2019MNRAS.486.3019F}.
We define the gravitationally bound stellar mass as the cluster mass. 
Because our snapshots are output every 0.025 or 0.05\,Myr because of the filesizes, the gas expulsion time is determined in 0.025 or 0.05\,Myr increments. 
Table~S3 lists the properties of the most massive clusters in each run; in most cases, the other clusters were less massive than $10^4 M_{\odot}$, except for model M1e6R10Z002v05s2. For this model, Table~S3 also lists the second most massive cluster, which was more than $10^4 M_{\odot}$.
The final cluster mass depends on the initial virial ratio: a smaller virial ratio resulted in the formation of more massive single clusters.

We calculate the mass of the VMS due to runaway collisions and the remnant IMBH mass using \textsc{SSE} \cite{2002MNRAS.329..897H,2020A&A...639A..41B}. Here, we adopt a mass- and metalicity-dependent mass-loss rate \cite{2018A&A...615A.119V} for VMSs. As an example, Figure~S4 shows the mass evolution of the VMS after gas expulsion for model M1e6R10Z002v01s2. This star loses $\sim 6000 M_{\odot}$ and finally leaves a $3400 M_{\odot}$ IMBH.

Figure~S3 also shows the simulated cluster mass compared to the masses of the IMBHs. The IMBH mass is roughly 3\,\% of the cluster mass for ($M_{\rm cluster}<10^5M_{\odot}$). For more massive clusters ($M_{\rm cluster}>10^5M_{\odot}$), the BH masses are a smaller fraction of the cluster mass due to the high mass-loss rate.  
Figure S5 shows the initial mass ($M_{\rm g}$) and the cluster ($M_{\rm cl}$) formed in each simulation. The cluster masses have a large scatter, but their distribution is similar to the results of previous star-cluster formation simulations for solar-metallicity clouds; $M_{\rm cl}\sim M_{\rm g}^{0.8}$ \cite{2018NatAs...2..725H}. 

We also analyze the rotation in a manner similar to that adopted in observations. In observations, the mean line-of-sight velocity normalized by the velocity dispersion is often used as an indicator of rotation \cite{2018MNRAS.481.2125B}. Our simulations record the six-dimensional data (positions and velocities) of each star. We, therefore, begin by identifying the rotation axis, by calculating the inertial momentum tensor, and then calculate the mean line-of-sight velocity ($V$). We adopt this velocity as the rotation velocity. The radial distribution of rotation velocity ($V(r)$) and the same quantity normalized by the line-of-sight velocity dispersion at each radius ($\sigma(r)$) for model M1e6R10Z002v01 is shown in Figure~S6. The rotation velocity rapidly increases as the radius increases and reaches a peak at around 0.03\,pc. Then, it gradually decreases with radius. Observationally, the peak velocity ($V_{\rm peak}$) is measured and used as an indicator of the rotation. For our simulations, we calculate the mean line-of-sight velocity at radii $>0.02$\,pc and normalize it by the one-dimensional velocity dispersion of the entire region. This value is almost equivalent to the peak velocity normalized by the line-of-sight velocity dispersion. 
The rotation of simulated clusters is summarized in Table~S3. 

Figure~S7 shows the relation between the initial rotation of the molecular cloud and the cluster rotation. Although the number of samples is limited, we do not find any tight correlation between the initial molecular cloud rotation and the final cluster rotation. The initial rotation of the parental molecular clouds could affect the mass of the most massive cluster in the system formed in the clouds. Figure~S8A shows that molecular clouds with a higher angular momentum tend to form a less massive star cluster relative to the total stellar mass. The star-formation efficiency decreases as the initial cloud rotation increases (Figure~S8B). In a molecular cloud with high angular momentum, several small clusters are formed rather than a single star cluster at the center (Figure~S9). When the multiple clusters host massive stars and are ionized, the small clusters become unbound, so they do not produce a massive cluster. 

We do not find any correlation between the cluster rotation and the rate of runaway collisions (Figure~S10). The mass increase rate is 3--5\,\% of the cluster mass increase rate. We infer this is because the rotation is not dominant in the central region of the clusters (Figure~S6). 

\section{Analytic estimation of IMBH mass}
Because our simulation parameter space is limited to small GCs due to the computational cost, we estimate the IMBH masses formed in larger GCs based on our simulation results and theoretical models. 
The mass evolution of VMSs is given by equation 1, which shows that metallicity affects the mass-loss rate due to the stellar wind. 
We assume that ${N(t)\langle m\rangle}/{t}=M_{\rm cl}(t)/t$ and that this equals the mass increase rate of the cluster due to star formation ($\dot{m}_{\rm SF}$). Then, we obtain
\begin{equation}
    \dot{m}_{\rm col} \sim 4\times 10^{-3} f_{\rm c} \ln \Lambda \dot{M}_{\rm SF}.\tag{S3} 
    \label{eq:factor}
\end{equation}
Assuming that the cluster mass is in the range $10^4$--$10^6 M_{\odot}$ and therefore $N\sim10^4$--$10^6$, we obtain $\dot{m}_{\rm col} \sim 0.03$--$0.05f_{\rm c} \dot{M}_{\rm SF}$. 
In our simulations, the mass-increase rate follows: $\dot{m}_{\rm col}\sim0.03$--$0.05\dot{M}_{\rm SF}$ (Figure~S10), which is consistent with the adopted value.

We re-evaluate the factor of four in equation 4 following the previous derivation \cite{2002ApJ...576..899P}, and the collision rate is:
\begin{equation}
    \dot{m}_{\rm col} = \dot{N}_{\rm col} \langle\delta m\rangle, \tag{S4}
    \label{eq:define}
\end{equation}
where $\dot{N}_{\rm col}$ and $\langle\delta m\rangle$ are the collision rate and the average stellar mass to collide, respectively. We estimate the collision rate based on the binary formation rate via three-body encounters as
\begin{equation}
    \dot{N}_{\rm col} \sim 10^{-3} f_{\rm c}\frac{N_{\rm cl}}{t_{\rm rlx}}, \tag{S5}
\end{equation}
where $f_{\rm c}$ is the collision efficiency of binaries, and $N_{\rm cl}$ and $t_{\rm rlx}$ are the number of cluster stars and the half-mass relaxation time of the cluster, respectively. We follow their assumption. 

The average mass increase per collision, $\langle\delta m\rangle$ is estimated using the dynamical timescale of massive stars to sink to the cluster center and adopt the radial orbital evolution of a star with $m$ in an isothermal sphere;
\begin{equation}
     r\frac{dr}{dt} = -0.428 \ln \Lambda\frac{Gm}{V_{\rm c}}
     = -0.302 \ln \Lambda\frac{Gm}{\sigma}, \tag{S6}
     \label{eq:friction}
\end{equation}
where $r$ is the distance of the star from the cluster center, $G$ is the gravitational constant, and $V_{\rm c}$ is the constant circular velocity ($V_{\rm c}=\sqrt{2}\sigma$) [\cite{2008gady.book.....B}, their equation 8.11]

We determine the half-mass relaxation time
\begin{equation}
    t_{\rm rlx} = \left(\frac{R^3}{GM_{\rm cl}}\right)\frac{N_{\rm cl}}{8 \ln \Lambda}, \tag{S7}
\end{equation}
where $R$ is the characteristic cluster (half-mass) radius \cite{1987degc.book.....S}. The relaxation time slightly changes between different initial conditions. 
If we assume a singular isothermal sphere, the relaxation  time is:
\begin{equation}
    t_{\rm rlx} = 2.1 \frac{\sigma r^2}{G \langle m\rangle \ln \Lambda} \tag{S8}\label{eq:t_rlx}
\end{equation}
[\cite{2008gady.book.....B}, their equation 7.107]. 

Integrating equation S6 and substituting in equation S8, we obtain the dynamical-friction in-spiral timescale as
\begin{equation}
    t_{\rm f} = 0.79 \frac{\langle m \rangle}{m_{\rm f}} t_{\rm rlx}, \tag{S9}
\end{equation}
where $m_{\rm f}$ is the stellar mass that sinks to the cluster center in $t_{\rm f}$. This equation has a lower scaling factor (0.79) than was found in previous work (3.3) [\cite{2002ApJ...576..899P}, their equation 4]. We ascribe this to a slightly different form of the relaxation time. A real globular cluster does not have a singular isothermal distribution. 

For the mass increase per collision, $\langle \delta m\rangle$, we adopt an empirical relation [\cite{2002ApJ...576..899P}, their equation 11]: 
\begin{equation}
    \langle \delta m\rangle = 4 m_{\rm f} \tag{S10} \label{eq:simon}
\end{equation}

The mean mass in the cluster center at $t_{\rm f}$ is estimated as 
\begin{equation}
    \langle m \rangle _{\rm cen} = \frac{\int _{m_{\rm f}}^{m_{\rm max}} m^{\alpha +1}dm}{\int _{m_{\rm f}}^{m_{\rm max}} m^{\alpha}dm}, \tag{S11}\label{eq:mass_ave}
\end{equation}
where $m_{\rm max}$ is the upper-mass limit of the mass function, and $\alpha$ is the slope of the mass function in the cluster core. We assume $\langle m\rangle_{\rm col} = \langle \delta m \rangle$.
If we assume that the mass function in the cluster core is composed of stars with $>m_{\rm f}$, and that $\alpha$ is the same as in the initial mass function. This leads to $\langle m \rangle_{\rm col}=3.3 m_{\rm f}$ for $\alpha = -2.3$, $m_{\rm max}=150M_{\odot}$, and $m_{\rm f}=1M_{\odot}$. This value is similar to that adopted in previous work\cite{2002ApJ...576..899P}. In our simulations, the mass function of stars that merged with the VMS had a slope much shallower than that of the initial mass function (Figure~3C). The average mass of merged stars in our simulations was $\sim 6M_{\odot}$. We also adopt $m_{\rm f}\sim 1M_{\odot}$ from the mass function of stars merged to the VMS (Fig.~3), which is depleted in low-mass stars. From these values, we obtain $\langle \delta m \rangle \simeq 6m_{\rm f}$. The power-low slope of the merged stars is $\alpha \sim -1.8$ (Fig.~3C). Using eq.~(\ref{eq:mass_ave}), we obtain $\langle m \rangle_{\rm col}\sim 7 m_{\rm f}$ for $m_{\rm f}=1M_{\odot}$. This different value indicates there is theoretical uncertainty in the scaling factor. 

Substituting equations S5 and S10 into equation S4, we obtain
\begin{equation}
    \dot{m}_{\rm col} \sim A \times 10^{-3}f_{\rm c}\frac{N_{\rm cl}\langle m\rangle}{t}, \tag{S12}
\end{equation}
where $A$ is a scaling factor between 1 and 10. 

Although the factor has an uncertainty, we assume that $\dot{m}_{\rm col}=0.03\dot{M}_{\rm SF}$. This star-formation rate is defined using the cluster mass ($M_{\rm cl}$) and formation time ($t_{\rm form}$) as $\dot{M}_{\rm SF}=M_{\rm cl}/t_{\rm form}$. 
The formation timescale ($t_{\rm form}$) is assumed to equal the collapse time (free-fall time, $t_{\rm ff}$) of the natal GMC and scales with the inverse square root of the gas density. To form a massive star, the gas density must be higher than the critical density, which is necessary for the gas to collapse, overcoming the feedback from the massive stars. The critical surface density is estimated to be $\sim 10^3M_{\odot}$\,pc$^{-2}$ \cite{2022MNRAS.511.3346F}, corresponding to $\sim10^2\,M_{\odot}$\,pc$^{-3}$ for a typical radius of massive GMC, $\sim 10$\,pc. The resulting $t_{\rm ff}$ is $\sim 0.8$\,Myr. We adopt this value as the maximum $t_{\rm ff}$. We set the minimum $t_{\rm ff}$ as 0.3\,Myr, corresponding to $\sim 700 M_{\odot}$\,pc$^{-3}$ which is lower than the threshold density for star formation. We adopt $t_{\rm form}= 2\,t_{\rm ff}$, following the results of a previous study \cite{2022MNRAS.511.3346F}. This free-fall time range results in the range in IMBH mass for a given cluster mass. 
To determine the cluster mass, we assume a relation between the initial gas mass and cluster mass: $M_{\rm cl} = M_{\rm g}^{0.8}$ \cite{2018NatAs...2..725H}. Thus, the cluster mass and its formation time are derived for a given initial gas mass and density.
We assume that the mass increase of VMSs and their host clusters stops at $t_{\rm form}$. After $t_{\rm form}$, the VMS loses mass owing to stellar winds. 

We integrate the mass of the VMS with $\dot{m}_{\rm col}$ and $\dot{m}_{\rm SW}$, assuming that the main-sequence lifetime is 2\,Myr\cite{2018MNRAS.478.2461G}. The evolution of VMSs with masses $>10^3M_{\odot}$ is unclear; it has been proposed that they evolve into red supergiants that lose their mass through a strong stellar wind, although the duration ($\sim 0.3$\,Myr) is shorter than the main-sequence lifetime \cite{2020ApJ...902...81N}. Following previous studies, we adopt a scaling relation for the mass-loss rate during the red-supergiant phase ($\dot{m}_{\rm RSG}$): $\log[\dot{m}_{\rm RSG}/M_{\odot}\,{\rm yr}^{-1}] = -2.88+ \log[m/10^3M_{\odot}]$\cite{2020ApJ...902...81N}. We also assume that all the remnant mass collapses into the IMBH \cite{2002RvMP...74.1015W}. 
With these, we obtain an analytic relation between cluster and IMBH masses shown as red and blue colored regions in Fig.~3. Because the free-fall time depends on the initial gas density, the results have a range in IMBH mass.

\section{The properties of Milky Way globular clusters}
Table~S4 lists the current mass ($M_{\rm GC, cur}$)\cite{2018MNRAS.478.1520B}, metallicity, power-law slope of the current stellar mass function of the cluster ($\alpha$) of Milky Way GCs\cite{1996AJ....112.1487H,2010arXiv1012.3224H}, which previous studies have suggested might contain IMBHs\cite{2010ApJ...710.1063V,2011A&A...533A..36L,2013A&A...554A..63F,2013A&A...552A..49L,2014A&A...566A..58K,2017Natur.542..203K,2019ApJ...875....1M,2020A&A...644A.133H,2021MNRAS.507.4788G}.   
All these GCs have ages $>10$\,Gyr and have lost a large fraction of their initial mass. To estimate the initial mass of the Milky Way GCs, we adopt a relation between $\alpha$ and the current-to-initial cluster mass ratio ($M_{\rm GC, cur}/M_{\rm GC,ini}$)\cite{2015MNRAS.453.3278W}:
\begin{eqnarray}
\alpha = A \times e^{B\frac{M_{\rm GC,cur}}{M_{\rm GC,ini}}}+C,
\end{eqnarray}
where $A$, $B$, and $C$ are fitting parameters obtained form previous simulations. We adopt $A=4.6$, $B=-7.2$, and $C=-2.2$ \cite{2015MNRAS.453.3278W}. The estimated initial masses are listed in Table~S4. They are typically ten times more massive than the current masses. 
Table S4 also lists the kinematically estimated IMBH masses plotted in Fig.~2. 

The existence and estimated masses of IMBHs in GCs are debated. Here, we discuss the proposed IMBH masses for some of the GCs in Table~S4, which have some disagreement among some independent estimations. 
$\omega$ Centauri ($\omega$ Cen) is one of the well-known candidates of IMBH-host GCs. It has been estimated that it hosts a 3--$5\times10^4M_{\odot}$ IMBH by comparing the velocity dispersions in the central region with a dynamical model \cite{2008ApJ...676.1008N,2010ApJ...719L..60N}. 
Another study \cite{2010ApJ...710.1063V} estimated an upper limit of $<1.2\times 10^4M_{\odot}$ for the IMBH in the same GC.
A similar IMBH mass, $\sim 4\times10^4 M_{\odot}$, has been estimated for $\omega$ Cen by comparison with $N$-body simulations \cite{2017MNRAS.464.2174B}.
Other studies \cite{2017MNRAS.468.4429Z,2019MNRAS.482.4713Z} have claimed that the observed velocity dispersions can be explained without an IMBH. X-ray observations have been interpreted as indicating an order of magnitude smaller IMBH masses $1000$--$5000 M_{\odot}$\cite{2011ApJ...729L..25L,2013ApJ...773L..31H}, but these could be due to a low accretion rate onto the IMBH. Thus, the estimated IMBH mass of $\omega$ Cen is uncertain. We adopt the middle mass ($<1.2\times 10^4M_{\odot}$) for $\omega$ Cen.

For 47 Tucanae (47 Tuc), it has been suggested that the maximum additional mass required at the center is $1000$--$1500M_{\odot}$ from the observed velocity dispersion, although that can be explained without an IMBH \cite{2006ApJS..166..249M}. Another study\cite{2017Natur.542..203K} estimated that 47 Tuc
hosts a $2300^{+1500}_{-850}M_{\odot}$ IMBH using a comparison between observed accelerations of millisecond pulsars and $N$-body simulations. We adopt this mass as the upper limit. 
However, other studies claimed that no IMBH is necessary for 47 Tuc\cite{2019ApJ...875....1M,2020A&A...644A.133H,2020ARA&A..58..257G}.

\begin{figure}[h]%
\centering
\includegraphics[width=0.98\textwidth]{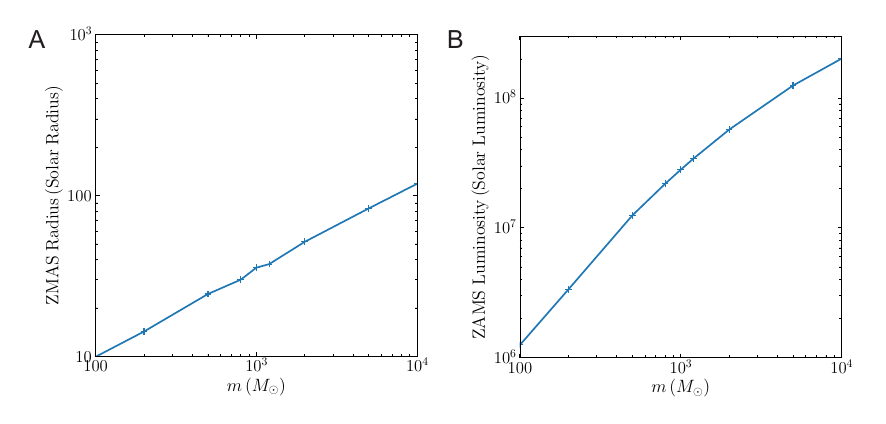}
\caption{\textbf{Zero-age main-sequence (ZAMS) radius (A) and luminosity (B) as a function of stellar mass ($m$).} The relations are  obtained using SSE for $Z=0.02Z_{\odot}$. 
}\label{fig:SSEini}
\end{figure}

\begin{figure}[h]%
\centering
\includegraphics[width=0.99\textwidth]{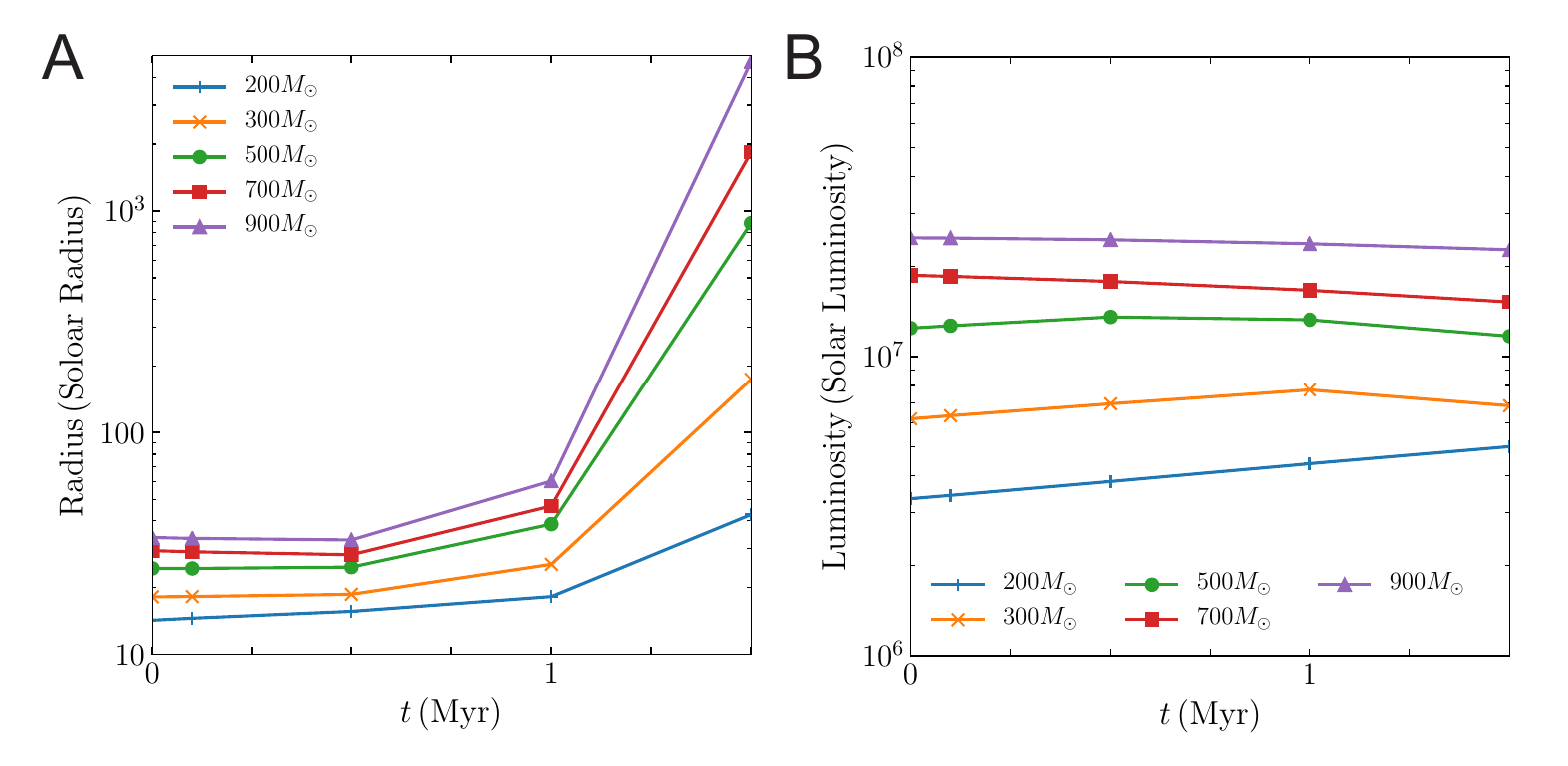}
\caption{\textbf{Radius (A) and luminosity (B) evolution of massive stars for $Z=0.02Z_{\odot}$ obtained using SSE. }Colors indicate the ZAMS masses from 200 to 900 $M_{\odot}$.
}\label{fig:SSEev}
\end{figure}

\begin{figure}[h]%
\centering
\includegraphics[width=0.9\textwidth]{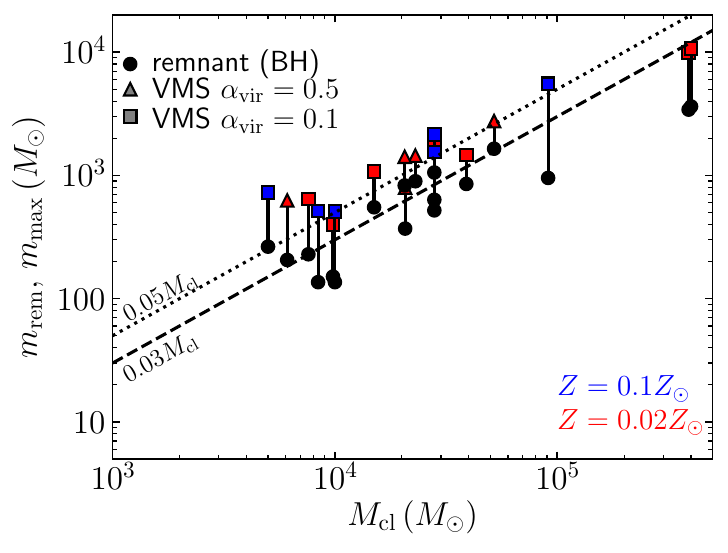}
\caption{\textbf{Relation between simulated cluster mass and maximum VMS (triangle and square) and remnant (circle) masses.} Colored symbols indicate the mass of the maximum mass of the VMSs. Triangles and squares indicate the initial virial ratios of 0.5 and 0.1, respectively. Cluster masses are measured in the last snapshot. Blue and red indicate $Z=0.1$ and $0.02 Z_{\odot}$, respectively. The dashed and dotted lines indicate $0.03M_{\rm cl}$ and $0.05M_{\rm cl}$, respectively.
}\label{fig:McMBH}
\end{figure}

\begin{figure}[h]%
\centering
\includegraphics[width=0.8\textwidth]{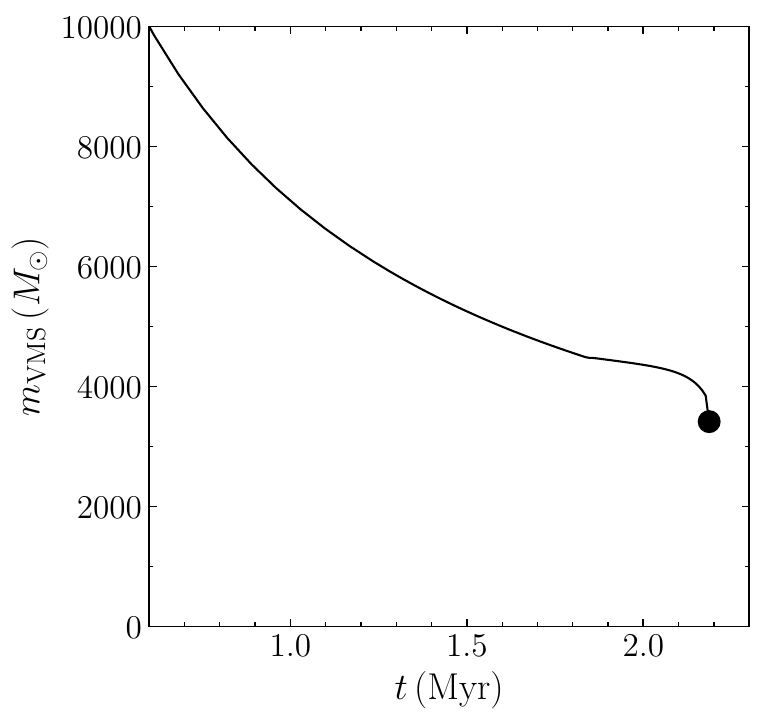}
\caption{\textbf{Mass evolution of the VMS after gas expulsion calculated for model M1e6R10Z002v01s2.} The black dot indicates the time and mass when a BH was formed. Here, $t$ indicates the simulation time and the $x$-axis starts from 0.6\,Myr. The mass evolution before 0.6\,Myr is shown in Fig.~1D.
}\label{fig:VMSevolution}
\end{figure}

\begin{figure}[h]%
\centering
\includegraphics[width=0.9\textwidth]{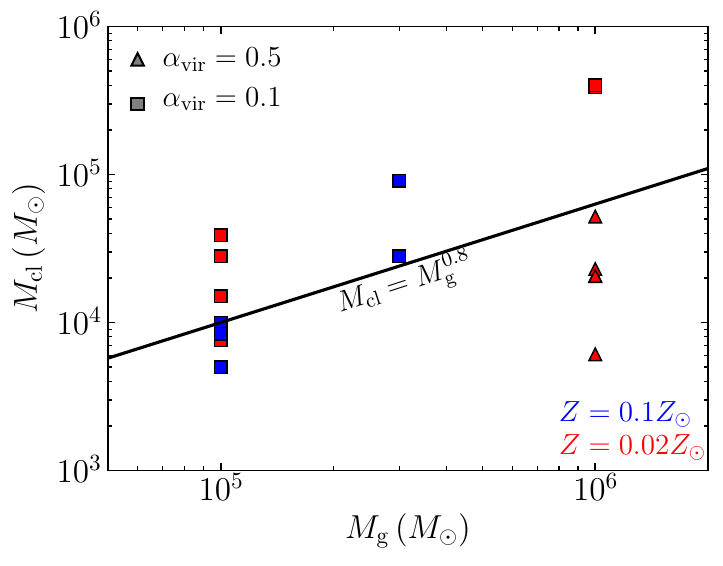}
\caption{\textbf{Relation between initial GMC mass and simulated cluster mass.} Triangle and square symbols indicate the virial ratio of the initial conditions. Colors are the same as Figure~S4. Black line indicates the relation obtained from previous simulations ({\it 28}).
}\label{fig:McMc}
\end{figure}

\begin{figure}[h]%
\centering
\includegraphics[width=0.99\textwidth]{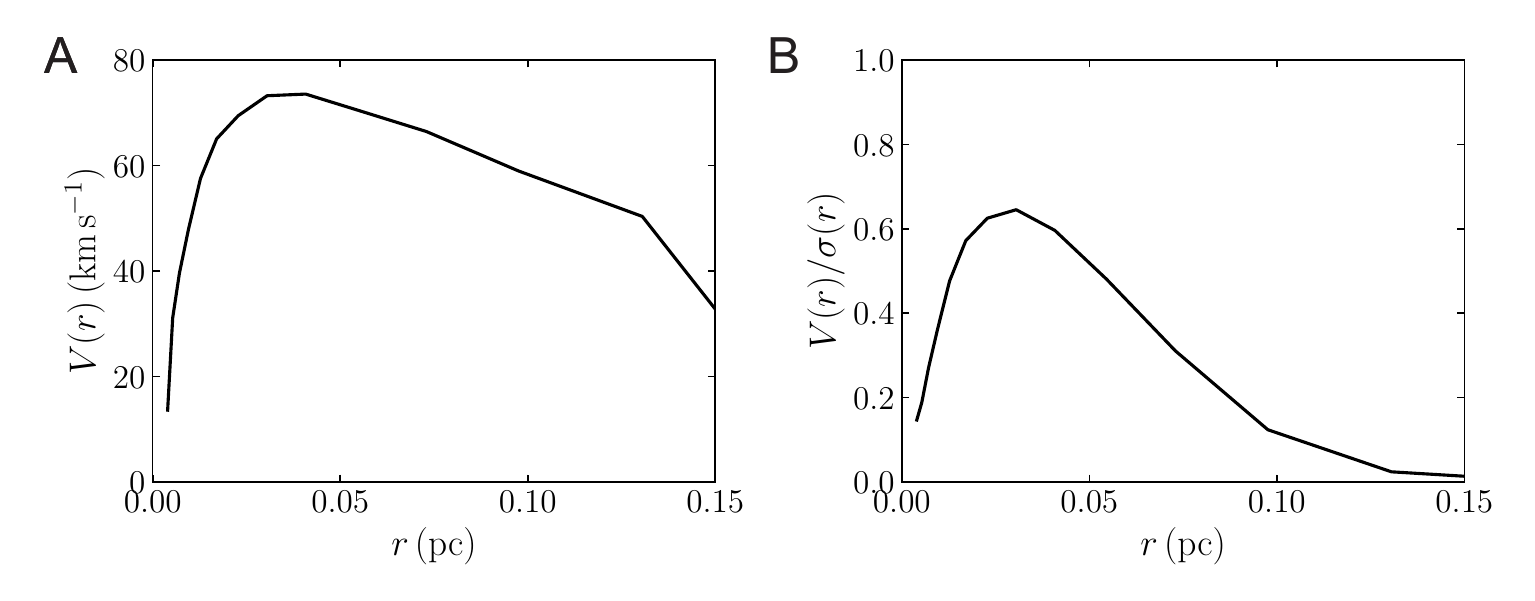}
\caption{{\bf An example of cluster rotation distribution.} Cluster rotation speed (A) and that scaled by the velocity dispersion (B) as a function of the distance from the cluster center for model M1e6R10Z002v01.} 
\label{fig:cluster_rot}
\end{figure}

\begin{figure}[h]%
\centering
\includegraphics[width=0.99\textwidth]{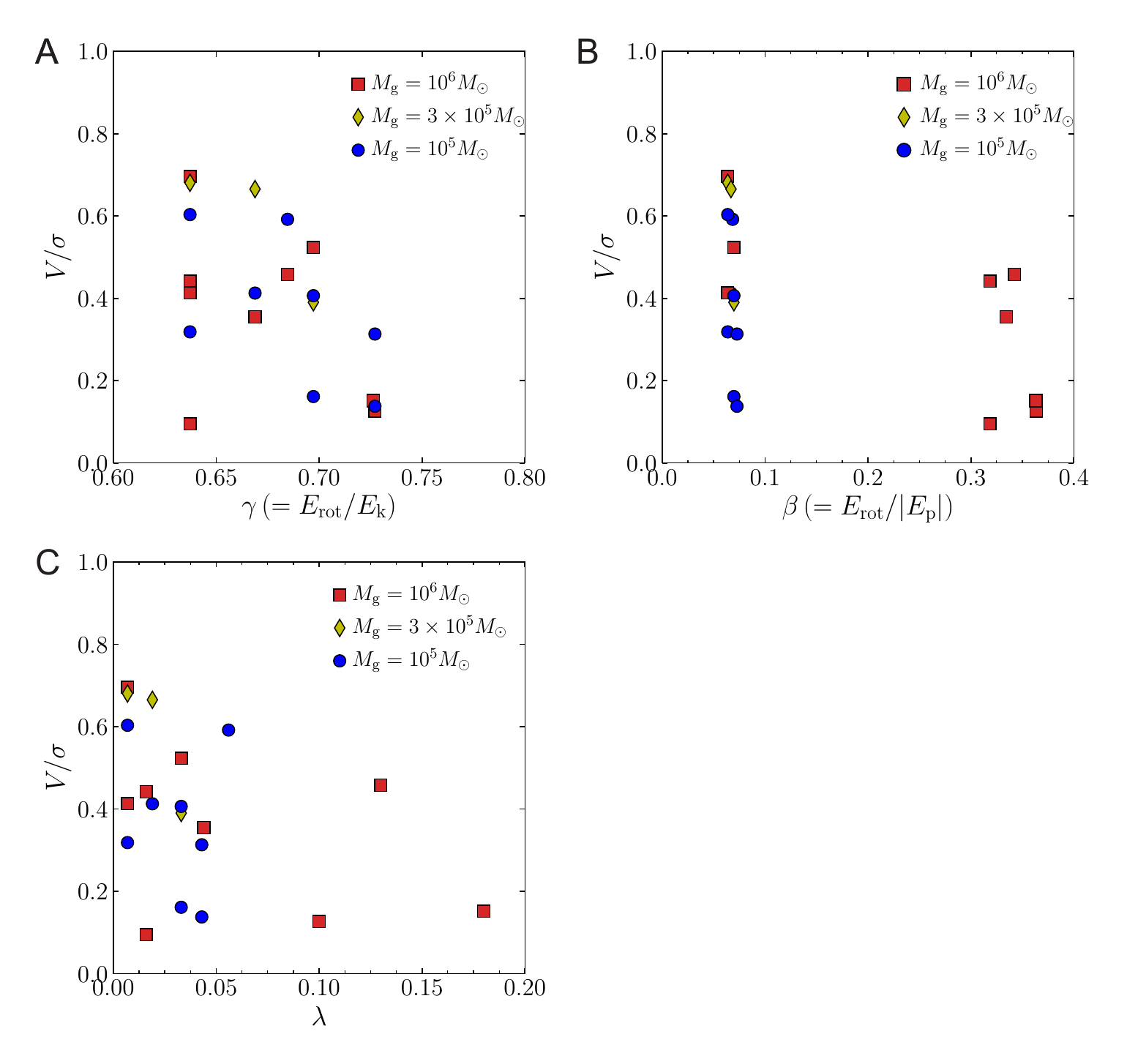}
\caption{{\bf The relation between initial cloud rotation and cluster rotation ($V/\sigma$).}  (A) to (C) are for $\gamma$, $\beta$, and $\lambda$, respectively. Red squares, yellow rhombuses, and blue circles indicate the initial cloud mass of $10^6$, $3\times 10^5$, and $10^5 M_{\odot}$.} 
\label{fig:IC_rot}
\end{figure}

\begin{figure}[h]%
\centering
\includegraphics[width=0.99\textwidth]{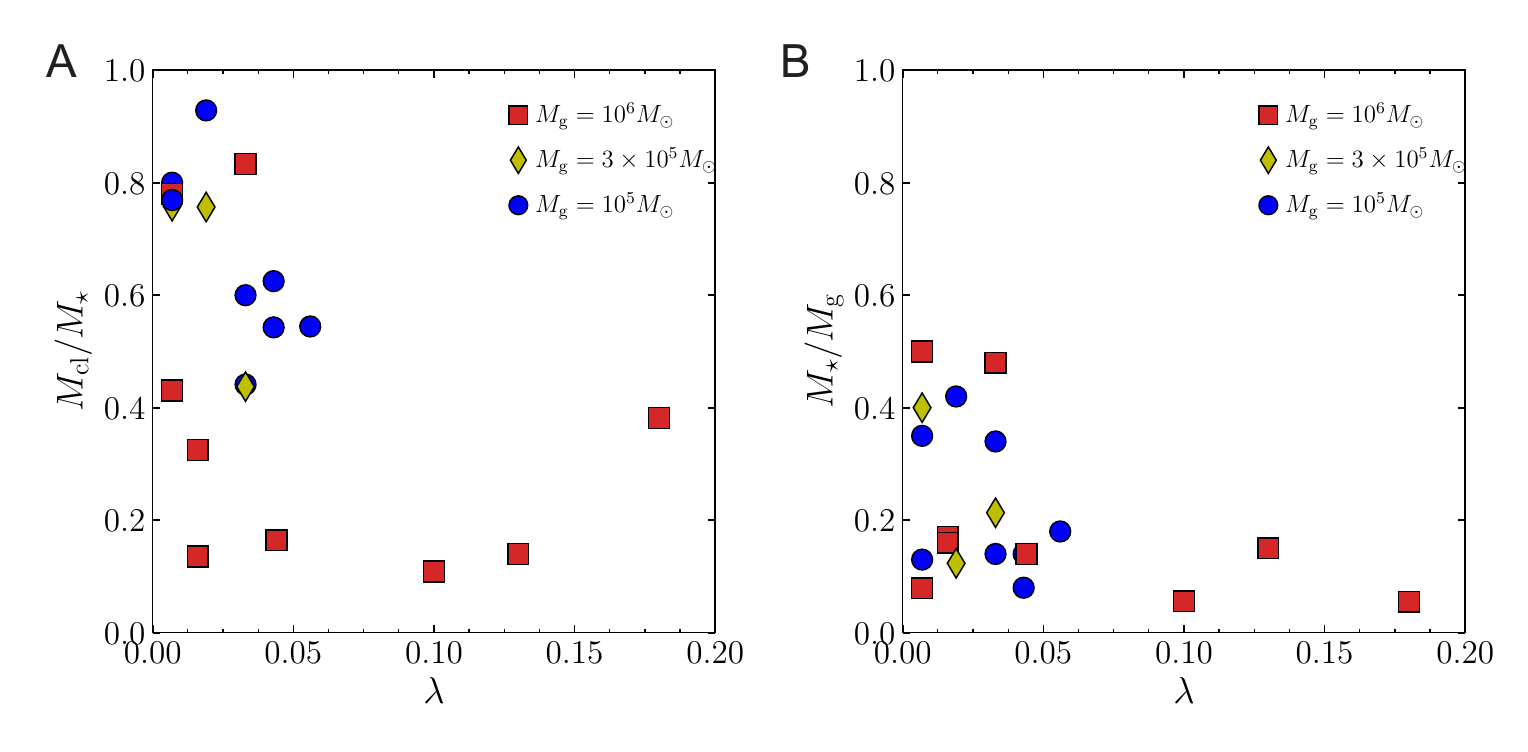}
\caption{{\bf The relation between initial cloud rotation ($\lambda$) and formed stellar mass.} A: the mass of the most massive clusters to the total stellar mass. B: formed stellar mass to the initial cloud mass (star-formation efficiency of the entire system). Symbols are the same as Figure~S7.} 
\label{fig:cluster_rot_mass}
\end{figure}

\begin{figure}[h]%
\centering
\includegraphics[width=0.99\textwidth]{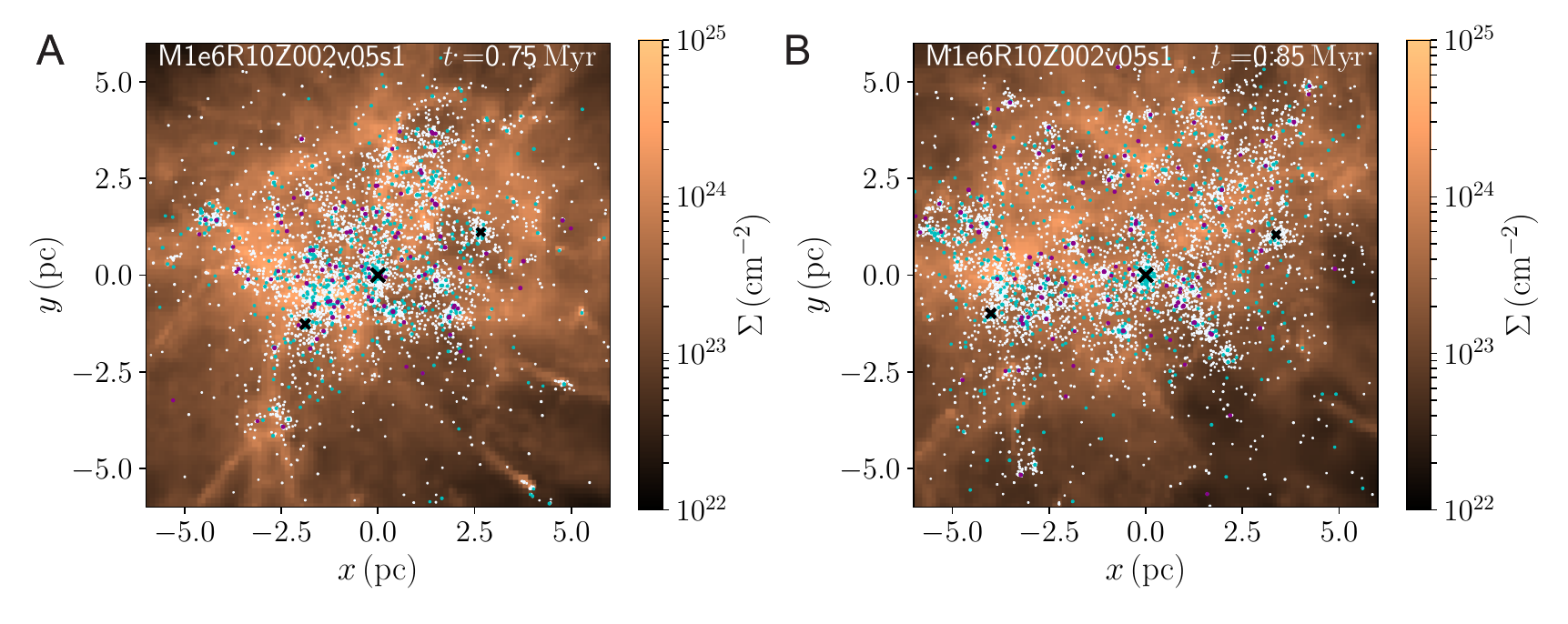}
\caption{{\bf Snapshots of model M1e6R10Z002v05s1.} A and B are for $t=0.75$ and 0.85 Myr, respectively. The color scale indicates the column density distribution of gas, and the dots indicate stars (purple: $m>40M_{\odot}$, blue: $10<m<40M_{\odot}$, white: $m<10M_{\odot}$). Black crosses indicate VMSs with $m>300M_{\odot}$, and the symbol size indicates the mass. } 
\label{fig:snap_M1e6v05}
\end{figure}

\begin{figure}[h]%
\centering
\includegraphics[width=0.8\textwidth]{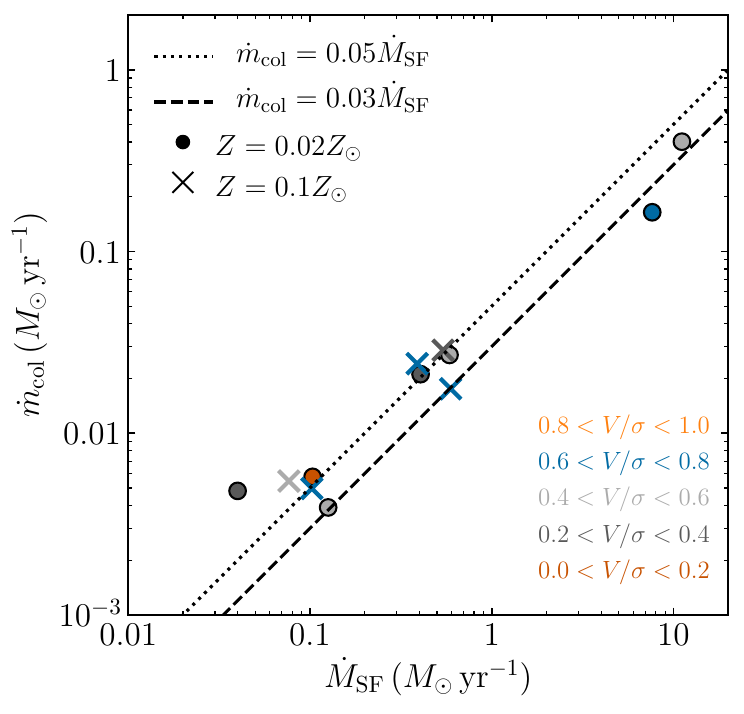}
\caption{{\bf Relation between the mass increase rates of the most massive star and cluster.} Colors indicate the rotation of clusters. We plot the models with $\alpha_{\rm vir}=0.1$ because the snapshot interval for the models with $\alpha_{\rm vir}=0.5$ was too long to measure the mass increase rates precisely.}
\label{fig:Mdot}
\end{figure}

\clearpage

\begin{table*}
\begin{center}
\caption{\textbf{Initial conditions\label{tb:IC}.} From the left: model name, metallicity ($Z$), initial cloud mass ($M_{\rm g}$), radius ($R_{\rm g}$), initial density ($n_{\rm ini}$), initial surface density ($\Sigma_{\rm ini}$), initial free-fall time ($t_{\rm ff, ini}$), virial ratio ($\alpha_{\rm vir}$), and the number of runs ($N_{\rm run}$).}
\begin{tabular}{lcccccccccc}
\hline
   Name  &$Z$ & $M_{\rm g}$ & $R_{\rm g}$& $n_{\rm ini}$ & $\Sigma_{\rm ini}$ & $t_{\rm ff,ini}$ & $\alpha_{\rm vir}$ & $N_{\rm run}$ \\
   
      & $(Z_{\odot})$ & $(M_{\odot})$ & (pc) &  (cm$^{-3}$) & ($M_{\odot}$\,pc$^{-2}$ ) & (Myr)   & &  \\ 
      \hline
  M1e5R5Z002v01 & 0.02 & $10^5$  & 5 & $5600$ &  $1300$ & $0.59$  & 0.1 & 5\\
  M1e6R10Z002v05 & 0.02 & $10^6$ & 10 & $7000$ & $3200$ & $0.52$ & 0.5 & 3 \\
  M1e6R10Z002v01 & 0.02 & $10^6$ & 10 & $7000$ & $3200$ & $0.52$ & 0.1 & 2\\
  M1e6R17Z002v01 & 0.02 & $10^6$ & 17 & $1400$ & $1100$ & $1.16$ & 0.1 & 1\\
 M1e5R5Z01v01 & 0.1 & $10^5$  & 5 & $5600$ &  $1300$ & $0.59$  & 0.1 & 3\\
 M3e5R7Z01v01 & 0.1 & $3\times10^5$  & 7.5 & $4500$ &  $1700$ & $0.62$  & 0.1 & 3\\
\hline
\end{tabular}
\end{center}
\end{table*}

\begin{table*}
\begin{center}
\caption{\textbf{Rotation parameters for each random seed of initial turbulent velocity field\label{tb:seed}.} The combination between the models given in Table~S\ref{tb:IC} and random seeds are given in Table~S\ref{tb:results}.}
\begin{tabular}{lcccc}
\hline
Random seed  & \multicolumn{2}{c}{$\lambda$} & \multicolumn{2}{c}{$\beta$}\\
   & $\alpha_{\rm vir}=0.5$ & $\alpha_{\rm vir}=0.1$ & $\alpha_{\rm vir}=0.5$ &$\alpha_{\rm vir}=0.1$ \\
      \hline
s1 & 0.13 & 0.056 & 0.34 & 0.068\\
s2 & 0.016 & 0.0069 & 0.32 & 0.064 \\
s3 &  0.10 & 0.043 & 0.36 & 0.073 \\
s4 & 0.071 & 0.033 & 0.35 & 0.070 \\
s5 & 0.044 & 0.019 & 0.33 & 0.067 \\
s6 & 0.18 & 0.081 & 0.36 & 0.073\\
\hline
\end{tabular}
\end{center}
\end{table*}

\begin{table*}
\begin{center}
\caption{\textbf{Summary of the simulation output.} The columns show the model name, gas expulsion time ($t_{\rm GE}$), bound cluster mass ($M_{\rm cl}$), half-mass radius ($r_{\rm h}$) and half-mass density ($\rho_{\rm h}$) of the cluster, total stellar mass of the entire region ($M_{\star}$), the mass of the most massive star ($m_{\rm max}$), the BH mass obtained from the most massive star ($m_{\rm rem}$), and the cluster rotation velocity ($V$) scaled by one-dimensional velocity dispersion ($\sigma$).} \label{tb:results}
\begin{tabular}{lcccccccc}
\hline
   Name  & $t_{\rm GE}$ &$M_{\rm cl}$ & $r_{\rm h}$ & $\rho_{\rm h}$ & $M_{\star}$ & $m_{\rm max}$  & $m_{\rm rem}$ & $V/\sigma$\\
  
      & (Myr) & ($10^4M_{\odot}$) & (pc) & ($10^4M_{\odot}$\,pc$^{-3}$)  & ($10^4M_{\odot}$) & ($M_{\odot}$) & ($M_{\odot}$) & \\ 
      \hline
  M1e5R5Z002v01s1 & 0.75 & 0.98 & 0.54 & 755 & 1.8 &  398 & 151 & 0.59 \\
  M1e5R5Z002v01s2 & 0.7 & 2.8 & 0.030 & $1.2\times 10^4$ & 3.5 &  1819 & 1055 & 0.32 \\ 
  M1e5R5Z002v01s3 & 0.75 & 0.76 & 0.049 & 784 & 1.4 & 645  &  229 & 0.31 \\ 
  M1e5R5Z002v01s4 & 0.725 & 1.5 & 0.029 & 7085 & 3.4 &  1072 & 551 & 0.16 \\ 
  M1e5R5Z002v01s5 & 0.7 & 3.9 & 0.026 & $2.7\times10^4$ & 4.2 & 1467 & 852 & 0.41 \\ 
  M1e6R10Z002v05s1 & 0.85 & 2.1 & 0.11 & 210 & 15 & 795 & 370 & 0.46 \\
  M1e6R10Z002v05s2-1 & 0.7 & 5.2 & 0.061 & 2607 & 17 & 2754 & 1645 & 0.44 \\ 
  M1e6R10Z002v05s2-2 & 0.65 & 2.3 & 0.012 & $1.7\times10^5$ & 16 & 2770 & 1485 & 0.10 \\ 
  M1e6R10Z002v05s3 & 0.6 & 0.61 & 0.023 & $4.0\times 10^4$ &  5.6 & 626 &  206 &  0.13 \\
  M1e6R10Z002v05s5 & 0.8 & 2.3 & 0.11 & 228 & 14 & 1448 & 896 & 0.35 \\
  M1e6R10Z002v05s6 & 0.55 & 2.1 & 0.057 & 1303 & 5.5 & 1416 &  828 &  0.15 \\
  
  M1e6R10Z002v01s2 & 0.6 & 39 & 0.025 & $2.9\times 10^5$ & 50 & 9896 & 3415 & 0.69 \\ 
  M1e6R10Z002v01s4 & 0.6 & 40 & 0.025 & $3.1\times10^5$ & 48 & 10714 & 3622 & 0.52 \\ 
  M1e6R17Z002v01s2 & 1.175 & 3.4 & 0.046 & 3761 & 7.9 & 1800 & 1029 & 0.41 \\
  \\
  M1e5R5Z01v01s2 & 0.75 & 1.0 & 0.031 & 4197 & 1.3 & 511 & 136 & 0.60 \\ 
  M1e5R5Z01v01s3 & 0.7 & 0.50 & 0.020 & 7736  & 0.80 & 726 & 264 & 0.14 \\ 
  M1e5R5Z01v01s4 & 0.75 & 0.84 & 0.036 & 2084 & 1.4 & 513 & 136 & 0.44\\ 
  M3e5R7Z01v01s2 & 0.85 & 9.1 & 0.020 & $1.4\times 10^5$ & 12 & 5541 & 953 & 0.68 \\ 
  M3e5R7Z01v01s4 & 0.725 & 2.8 & 0.021 & $3.6\times10^4$ & 6.4 & 2145 & 634 & 0.39 \\ 
  M3e5R7Z01v01s5 & 0.75 & 2.8 & 0.035 & 7767 & 3.7 & 1546 & 520 & 0.66 \\ 
\hline
\end{tabular}
\end{center} 
\end{table*}

\begin{table*}[h]
\begin{center}
\caption{{\bf Globular cluster properties\label{tb:GC}.} Listed are cluster name, metallicity ({\it 48, 49}), the power-law slope of the measured mass function ({\it 48, 49}), the current mass of the cluster ({\it 33}), the estimated initial mass of the cluster, observationally estimated IMBH mass, and references for IMBH mass.}
\begin{tabular}{lcccccc}
\hline
   Name  & $\log (Z/Z_{\odot})$ & $\alpha$ & $M_{\rm GC, cur}$ & $M_{\rm GC, ini}$& $m_{\rm IMBH}$  & Ref.\\
      &  & & $(\log_{10}M_{\odot})$ & $(\log_{10}M_{\odot})$ & $(10^3M_{\odot})$  & \\ 
      \hline
47Tuc (NGC 104) & $-$0.72 & $-$0.53 & 5.89 & 6.74 & $<2.3\pm1.0$ & ({\it 85--87})\\  
NGC 1851 & $-$1.18 &$-$0.54& 5.48 & 6.32 & $<2.0$& ({\it 16})\\
NGC 1904 & $-$1.60 & $-$0.4 & 5.23 & 6.11 & $3\pm1$ & ({\it 16})\\
$\omega$ Cen (NGC 5139) & $-$1.53 & 0.0 & 6.55 & 7.54 & $<12$ & ({\it 81})\\
NGC 5286 & $-$1.69 & $-$0.67 & 5.60 & 6.42 & $3.9\pm2.0$& ({\it 83})\\
NGC 5694 & $-$1.98 &$-$0.8 & 5.58 & 6.37 & $<8.0$ &  ({\it 16}) \\
NGC 5824 & $-$1.91 & $-$1.0 & 5.89 & 6.62 & $<6.0$ & ({\it 16})\\
M80 (NGC 6093) & $-$1.75 & $-$0.31 & 5.40 & 6.30 & $4.6\pm1.4$ & ({\it 88})\\
NGC 6266 & $-$1.18 & $-$0.43 & 5.85 & 6.73 & $2.0\pm1.0$ &  ({\it 16})\\
NGC 6388 & $-$0.55 & $-$0.57 & 6.03 & 6.87 & $<2.0$  &  ({\it 82})\\
M3 (NGC 5272) &$-$1.50 & $-$0.75 & 5.60 & 6.39 & $<5.3$ & ({\it 84})\\ 
M13 (NGC 6205) & $-$1.53 & $-$0.62 & 5.66 & 6.48 & $<8.1$ & ({\it 84})\\
M92 (NGC 6341) & $-$2.31 & $-$0.82 & 5.43 & 6.20 & $<0.98$ & ({\it 84})\\
\hline
\end{tabular}
\end{center}

\end{table*}

\end{document}